\documentclass[11pt,onecolumn]{article}
\usepackage{graphicx}% Include figure files
\usepackage{dcolumn}% Align table columns on decimal point
\usepackage{bm}% bold math

\usepackage{color}

\newcommand{\gammadot}{\dot \gamma}

\usepackage[margin=2.5cm]{geometry}
\usepackage{amssymb}
\usepackage{abstract}

\usepackage[authoryear, round]{natbib}

\title{The Jamming Perspective on Wet Foams}

\author{ \large Gijs Katgert \\
\footnotesize School of Physics \& Astronomy and SUPA, The University of Edinburgh, \\
\footnotesize The Kings Buildings, Mayfield Road, Edinburgh EH9 3JZ, United Kingdom. \\ 
\and \\
\large Brian P. Tighe \\
\footnotesize  Instituut-Lorentz, Universiteit Leiden, 
\footnotesize  Postbus 9506, 2300 RA Leiden, The Netherlands. \\ 
\and \\
\large Martin van Hecke \\
\footnotesize Kamerlingh Onnes Laboratorium, Universiteit Leiden, 
\footnotesize  Postbus 9504, 2300 RA Leiden, The Netherlands. \\
\footnotesize mvhecke@physics.leidenuniv.nl}

\date{}
\begin{document}
\bibliographystyle{plainnat}

\maketitle

\begin{abstract}

Amorphous materials as diverse as foams, emulsions, colloidal
suspensions and granular media can {\em jam} into a rigid,
disordered state where they withstand finite shear stresses before
yielding. The jamming transition has been studied extensively, in
particular in computer simulations of frictionless, soft, purely
repulsive spheres. Foams and emulsions are the closest realizations
of this model, and in foams, the (un)jamming point corresponds to
the wet limit, where the bubbles become spherical and just form
contacts. Here we sketch the relevance of the jamming perspective
for the geometry and flow of foams --- and also discuss the impact
that foams studies may have on theoretical studies on jamming.

We first briefly review insights into the crucial role of disorder
in these systems, culminating in the breakdown of
the affine assumption that underlies the rich mechanics near
jamming. Second, we discuss how crucial theoretical predictions,
such as the square root scaling of contact number with packing
fraction, and the nontrivial role of disorder and fluctuations for
flow have been observed in experiments on 2D foams. Third, we
discuss a scaling model for the rheology of disordered media that
appears to capture the key features of the flow of foams,
emulsions and soft colloidal suspensions. Finally, we discuss how
best to confront predictions of this model with experimental
data.

\end{abstract}

%\pacs{83.80.Iz,64.70.D-,83.00.00,61.43.-j}

% 83.00.00 Rheology
% 83.60.La Viscoplasticity; yield stress
% 61.43.-j Disordered solids  (structure) (see also  81.05.Kf Glasses,)
% 64.00.00 Equations of state, phase equilibria, and phase transitions
% 64.60.Ht Dynamic critical phenomena
% 64.70.D- Solid-liquid transitions
% 64.70.dm General theory of the solid-liquid transition
% 83.80.Fg Granular solids
% 83.80.Hj Suspensions, dispersions, pastes, slurries, colloids
% 83.80.Iz Emulsions and foams

\section{Introduction}

Amorphous materials as diverse as foams, emulsions, colloidal
suspensions and granular media can {\em jam} into a rigid,
disordered state where they withstand finite shear stresses before
yielding [\citet{jamcool,epitome, reviewnagel,jammingreview}]. The
jamming transition has been studied extensively, in particular in
computer simulations of frictionless, soft, purely repulsive
spheres. Foams and emulsions are the closest realizations of this
model, and in foams, the (un)jamming point corresponds to the wet
limit, where the bubbles become spherical and just touch. What
does jamming tell us about foams? And what do experiments on foams
teach us about jamming?

Liquid foams and emulsions are dispersions of gas bubbles or
droplets in a immiscible (second) liquid phase, stabilized by
surfactants. Phenomenologically, these materials exhibit plastic
flow under large stresses, but behave elastically under small
stress - their macroscopic response to mechanical perturbations is
a complex mix of elastic, plastic and viscous effects
[\citet{hohlerreview}] - as is the case for many other disordered
soft materials ranging from colloidal suspensions to granular
materials [\citet{SMrev}].

How should we think about the mechanics of such materials, and
more specifically, about the mechanics of foams? More precisely,
how are the laws that govern the local dynamics of bubbles and
soap films related to the macroscopic, collective behavior of a
foam?  It is useful to introduce two extreme points of view here
first, being fully aware that most researchers' views are more
subtle.

One extreme point of view would be that a complete knowledge of
the local mechanics is sufficient to capture the global behavior -
from this perspective, the translation from local to global
behavior is nothing but a straightforward coarse-graining
procedure, and nontrivial collective behavior, such as nonlinear
rheology, always finds its root cause in similar nontrivial local
events, such as a nonlinear viscous friction law. For reasons that
will become clear, we refer to this as the `affine' picture.

Another extreme point of view would be that none of the local
details matter and that the global behavior of disordered media,
and foams in particular, is set by ``universal'' collective
mechanisms [\citet{SGR,olssonprl07, argon, falk}]. Hence knowledge of the
detailed interactions is irrelevant, and different local
interactions, such as different local friction laws, may lead to
the same macroscopic behavior, i.e. rheology.

Here we put forward a third point of view: qualitatively new
behavior can emerge at the global level, but local details still
may matter. So, in our view, there is no universality in the sense
that changes in the microscopics often lead to changes in the
macroscopic properties. However, due to the general `non-affine'
and strongly fluctuating nature of the mechanics of disordered
systems, the translation from the local to the global level can be
also highly nontrivial. We suggest that the {\em mechanisms} that
connect the micro and macro world are universal. As an example of such a
robust mechanism, we will discuss the role of the strength of the
dynamic fluctuations, which in turn depend on the wetness of the foam
and the flow rate.

We will illustrate our point of view by discussing recent
experimental and numerical work on foams, and by making explicit
contact with the jamming framework that has been emerging in
recent years. Jamming refers to the creation (or loss) of rigidity
in disordered systems in general, but the most studied models in
jamming are close to models for wet foams and
actually were inspired by earlier foam work [\citet{bolton, durianprl95}].

The outline of this paper is as follows. In section 2 we discuss
the basic jamming scenario for foams, including the differences between the mechanics of  disordered foams and
ordered models --- we conclude that disorder leads to strong
non-affinity, which becomes increasingly dominant near jamming. In
section 3 we discuss our recent observations of the nontrivial
scaling law that relates contact number and packing fraction of
foams and jammed materials --- this scaling law was first seen in
simulations twenty years ago [\citet{bolton}], and has now finally been observed
in 2D foams [\citet{katgertepl}]. Section 4 deals with our recent observations of the
difference in rheology between ordered and disordered foams, and
evidences both a nontrivial rheology for disordered 2D foams and a nontrivial scaling of the dynamic fluctuations [\citet{katgertprl, mobiusepl}]. In
section 5 we sketch the outlines of a model for the rheology of
disordered media near jamming that we recently introduced [\citet{tigheprl10}] which reproduces the experimental observations discussed in
section 4 well. In section 6 we briefly outline how best to confront our predictions with experimental data, and we close with a
short discussion and outlook.

\section{Jamming and Non-affinity: Consequences of Disorder}

{\bf Jamming Scenario for Foams ---} Some of the earliest studies
that consider the question of rigidity of packings of particles
concern the loss of rigidity in foams and emulsions with
increasing wetness [\citet{princen1983, princenIII, kraynik}]. The gas fraction $\phi$ plays a
crucial role in determining the foam's structure and rigidity. The
interactions between bubbles are repulsive and viscous, and static
foams are similar to the frictionless soft spheres used in most
models for jamming. In real foams, gravity (which causes drainage)
and gas diffusion (which causes coarsening) may play a role,
although these effects can be minimized by studying quasi-2D foams  
and using inert gases [\citet{weairebook}].

The (un)jamming scenario for foams is illustrated in Fig.~1. When
the gas fraction approaches one, the foam is called dry.
Application of deformations causes the liquid films to be
stretched, and the increase in surface area then provides a
restoring force: dry foams are jammed. When the gas fraction is
lowered and the foam becomes wetter, the gas bubbles become
increasingly spherical, and the foam loses rigidity for some
critical gas fraction $\phi_c$ where the bubbles lose contact
(Fig.~1). The unjamming transition is thus governed by the gas
fraction, which typically is seen as a material parameter. For
emulsions the same scenario arises [\citet{masonprl95}].

\begin{figure}[tbh]
\begin{center}
\includegraphics[width=.48\textwidth]{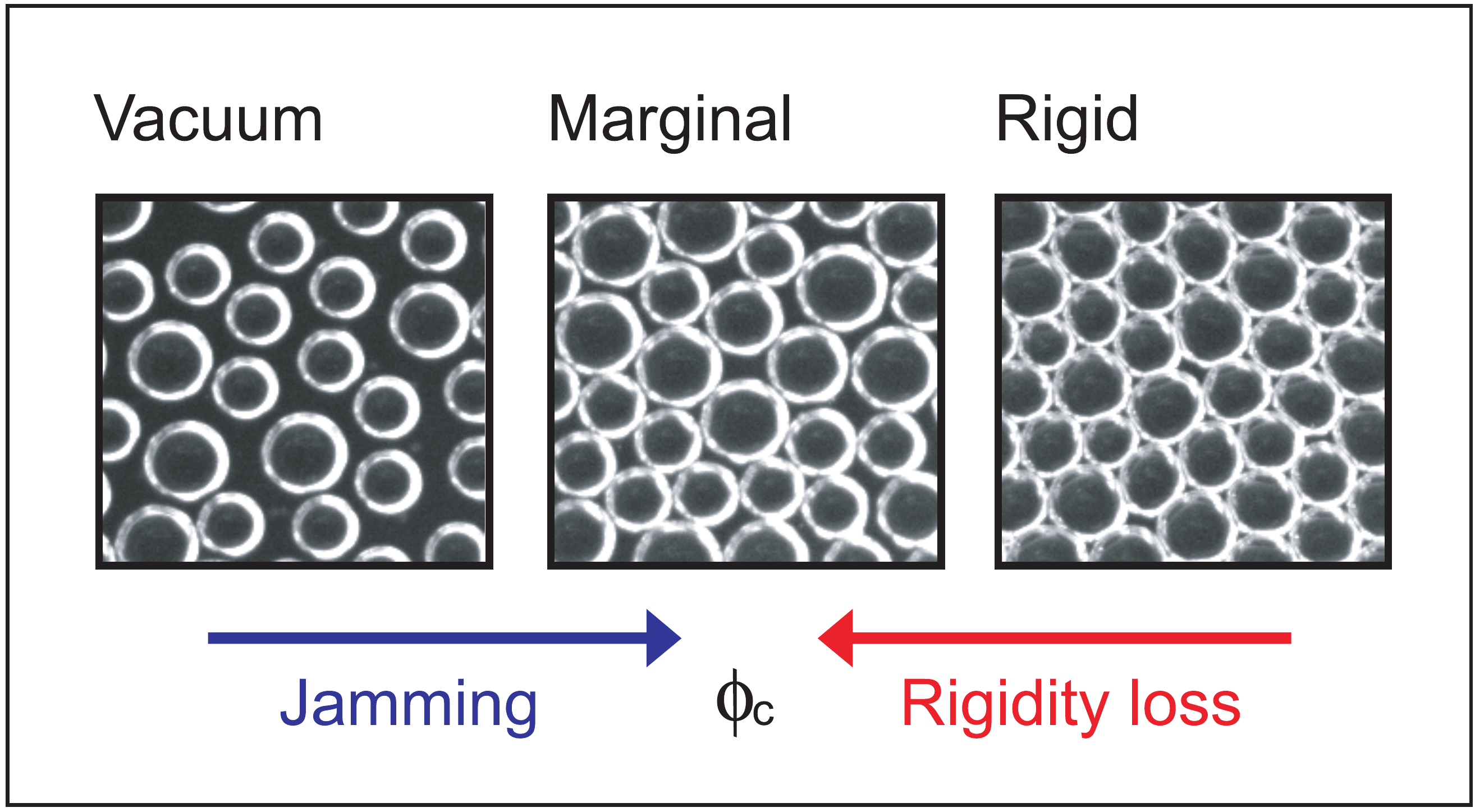}
\caption{Topviews of 2D foams, consisting of a mix of 2 and 3 mm
bubbles trapped below a top plate. At low packing fractions
(left), the bubbles do not form contacts and the materials is in a
mechanical vacuum state. At high packing fractions (right), the
bubbles are squeezed together and form a jammed, rigid state. At
intermediate packing fractions, the bubbles just touch and form a
marginal state. } \label{fig1}
\end{center}
\end{figure}

{\bf Elasticity of Disordered Media ---} Foams and emulsions typically form highly disordered packings. Is
this disorder important for the mechanical properties, and if so,
how can one deal with it?

The simplest approach is to ignore disorder altogether and start
from ordered, ``crystalline'' packings. Tacitly assuming  that all
bubbles experience the same deformation, analytical calculations
are then feasible, because one only needs to consider a single
bubble and its neighbors to capture the foams' geometry and
mechanical response.

A famous example are two-dimensional hexagonal packings of
monodisperse bubbles (``liquid honeycombs'' [\citet{princen1983, kraynik}]). The
only control parameter is then the liquid fraction of the foam or,
alternatively, the packing fraction of the gas bubbles $\phi$ (the
sum of liquid fraction and packing fraction is one).

The most striking features of these liquid honeycombs are as
follows. {\em{(i)}} The bubbles lose contact at the critical
density $\phi_c$ equal to $\frac{\pi}{2\sqrt{3}} \approx0.9069$,
and ordered foam packings are jammed for larger densities. {\em{(ii)}} When for such a model foam
$\phi$ is lowered towards $\phi_c$, the yield stress and shear
modulus remain finite and of similar order, and jump to zero
precisely at $\phi_c$. {\em{(iii)}} The
contact number (average number of contacting neighbors per bubble)
remains constant at 6 in the jammed regime. Similar results can be
obtained for three-dimensional ordered foams, where $\phi_c$ is
given by the packing density of the HCP lattice
$\frac{\pi}{3\sqrt{2}} \approx 0.7405$ [\citet{hoehlerosm}].

Confronted with experiments on  foams and emulsions, or numerical experiments
on disordered foams, all three of these predictions fail! First,
the critical packing fraction is substantially lower, around 84\%
in 2D and 64\% in 3D [\citet{bolton,durianprl95,masonprl95, stjalmes, lespiat}]. The fact that the critical packing density for ordered systems is higher
than that for disordered systems may not be a surprise, given that the bubbles are undeformed spheres at the jamming threshold, and
it is well known that ordered sphere packings are denser than
irregular ones [\citet{davescience}].

Second, the yield stress and shear modulus vanish smoothly when
the critical packing fraction is reached. Early evidence comes
from measurements for polydisperse emulsions by \citet{princenIII}, who observed a substantial lowering of
the shear modulus when $\phi$ is lowered. Later measurements by \citet{masonprl95,masonjcis96, masonpre97} and \citet{stjalmes} of the shear modulus and osmotic pressure of
compressed disordered emulsions and foams found similar behavior for the loss of rigidity. When scaled appropriately with the Laplace pressure, which sets the local ``stiffness'' of the droplets,  the shear modulus grows continuously with $\phi$ and vanishes at $\phi_c \approx 0.635$, which corresponds to random close
packing in three dimensions. There is also ample numerical
evidence for this [\citet{bolton,durianprl95,epitome,jammingreview}] - we will come back to this below.

Third, the contact number was found to vary smoothly with packing
fraction. Early evidence comes from numerical simulations by \citet{bolton}, who numerically probed how a {\em
disordered} foam loses rigidity when its gas fraction is decreased
. It was found that the contact number $z$ decreases
smoothly with $\phi$. At $\phi=1$ the contact number equals six,
as expected. When $\phi \rightarrow \phi_c$, the contact number
appears to reach the {\em marginal} value, $z_c =4$. In related work
on the so-called bubble model developed for wet foams in 1995,
Durian reached similar conclusions for two-dimensional model
foams, and found that the contact number indeed approaches
$4(=2D)$ near jamming, and observed the nontrivial square root
scaling of $z-4$ with excess density for the first time. All these findings are consistent with
what is found in closely related models of frictionless soft
spheres near jamming [\citet{epitome, jammingreview}].

We thus conclude that disorder plays an absolutely crucial role,
and in general cannot be treated as a perturbation from the regular,
ordered case.

\begin{figure}[tbh]
\begin{center}
\includegraphics[width=.48\textwidth]{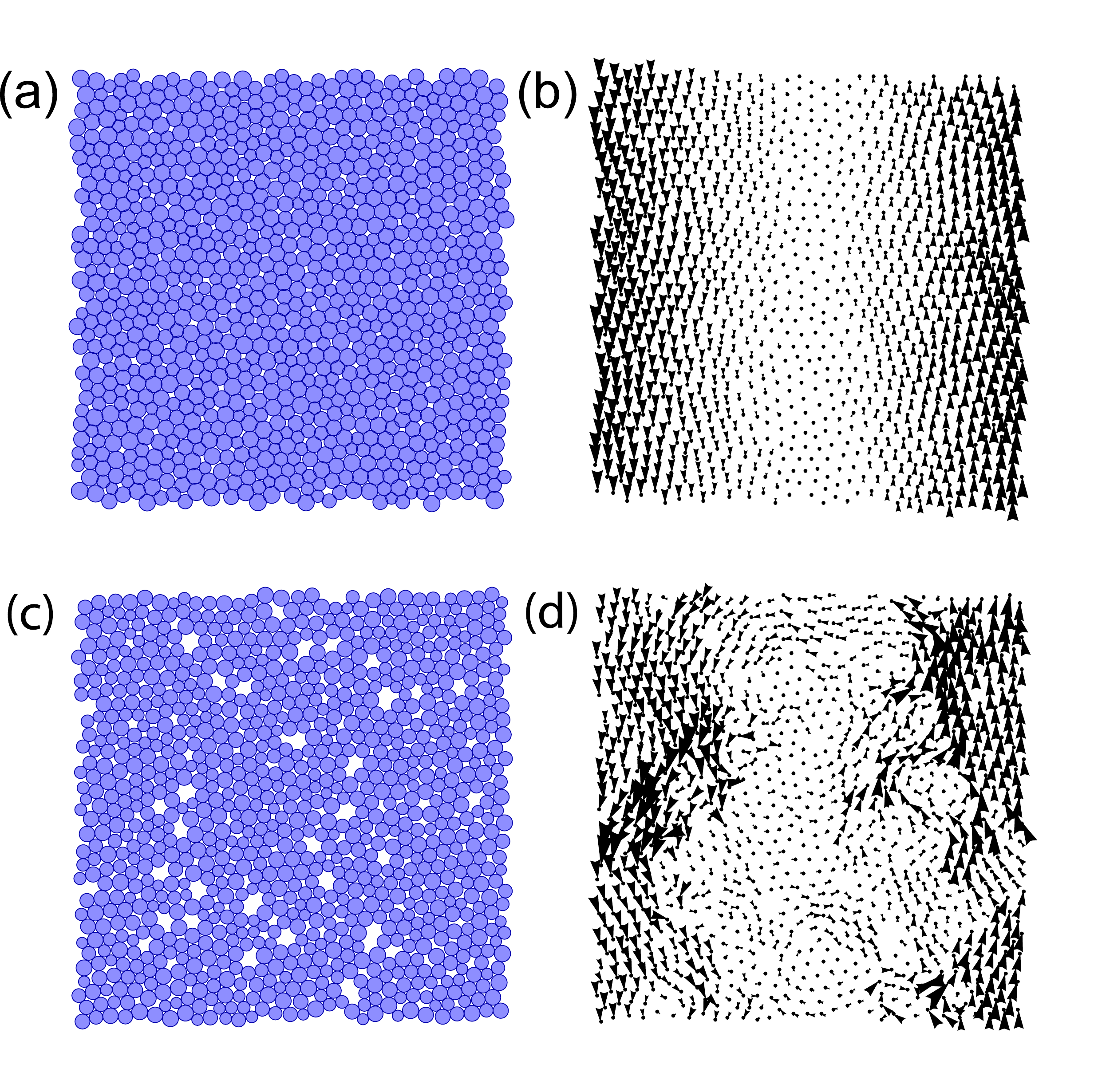}
\caption{(a) Dense numerical bubble packing, (b) Shear response of packing depicted in (a): the bubbles move affinely and thus largely follow the imposed strain. (c) Bubble packing close to jamming and its shear response (d): The bubbles exhibit non-affine motion and swirly flow patterns. } \label{fig2}
\end{center}
\end{figure}

{\bf Non-affinity ---} In the preceding subsection we presented a
wealth of simulational and experimental evidence that invalidates
simple predictions for the elasticity of disordered media based on
intuition derived from ordered packings. The crucial ingredient that is missing is the non-affine nature of the deformations of disordered
packings (Fig.~\ref{fig2}). In an affine deformation, all
the local deformations simply follow the global deformation,
implying that the local motion of the particles is as if they
where pinned to a rubber sheet. In a disordered system such as a
foam, where particles only interact with their neighbors, the
local, disordered environment is crucial. The key observation
is that, while materials far away from the jamming point (Fig.~2a)
exhibit deformations that are close to affine (Fig.~2b), materials
closer to the jamming point (Fig.~2c) exhibit deformations that
become increasingly disordered and non-affine (Fig.~2d).

The role of these non-affinities is to critically change the
elastic shear modulus of jammed materials such as foams
[\citet{epitome}]. The scenario is as follows: far away from
jamming, the local deformations are similar to the global ones -
so a 1\% shear strain leads to a typical change of the local deformations that is also of the order of 1\%. Since these changes of deformations are associated
with changes in the forces between bubbles/particles, and
concomittant changes in the elastic energy, one can then
immediately estimate the change in elastic energy once the
interactions between particles are known. In particular, for linear
interactions with a local spring constant $k$, one concludes that
the shear modulus $G$ on the order of $k$, just as in the
affine picture sketched above.

Closer to jamming, however, the number of contacts per bubbles
drops, and bubbles are increasingly free to deviate from the
affine field so as to minimize the changes in  elastic energy, and,
as a consequence, the shear modulus $G$ gets smaller than would be expected
from an affine assumption. At the jamming point, the shear modulus
vanishes for harmonically interacting particles --- clearly here
the translation from local to global interaction is nontrivial!
For different (e.g., power law interactions such a Hertzian
interactions), the shear modulus behaves different, so there is no
simple universality. However, the ratio of shear modulus to local
spring constant (or equivalently, the ratio of shear modulus to
bulk modulus $K$) scales in a robust manner with distance to
jamming: independent of dimension and interaction potential, $G/K
\propto z-z_c$, where $z_c$ is the number of contacts at the
critical point (=2D). [\citet{epitome, ellenbroekepl09}]. This illustrates our earlier
point: local interactions matter, but there are universal and
sometimes nontrivial mechanisms that translate the local
interactions to the global behavior.

As there is no simple way to estimate the particle motions and
deformations in disordered systems, one needs to resort to
(numerical) experiments. Jamming can be seen as the avenue that
connects the results of such experiments. Jamming aims at
capturing the mechanical and geometric properties of disordered
systems, building on two insights: first, that the non-affine
character becomes dominant near the jamming transition, and second,
that disorder and non-affinity are not weak perturbations away
from the ordered, affine case, but may lead to completely new
physics [\citet{ellakpre2005,masonprl95,granulence,tanguy,tanguy04,
lemaitre06,maloneyprl06}].

\section{Contacts and density in 2D foams near jamming}
We have recently experimentally investigated the static structure
of a disordered monolayer of foam bubbles that form a bidisperse
packing similar to the detail shown in Fig.~3. This monolayer
floats on the surface of a soapy solution and is bound on the top
by a well-leveled glass plate [\citet{katgertprl, katgertepl}].
This setup allows for direct optical access of all
bubbles, in contrast to three dimensional foams, which are opaque
due to multiple scattering of the interfaces. The
packing fraction can be varied simply by in- or decreasing the gap
between the glass plate and the soapy solution.

The question we set out to answer is whether the numerically
ubiquitous scaling of the excess contact number with the
square root of the excess density [\citet{durianprl95, epitome}] could be observed in
experiment. In order to probe this question we prepare many
distinct packings of our bidisperse foam at fixed $\phi$, by
stirring the bulk soapy solution to rearrange the packing. After
the packings have relaxed to an equilibrium state, we take
photographs of the resulting structure with a 6 megapixel camera
and analyse these by advanced image analysis. Even though the
bubble are three dimensional entities, we adjust the lighting of
our experiment such that we image the bubbles as two-dimensional
discs, at the point where they are broadest. This approach seems
to be justified, {\em a posteriori}. For each realization we determine the average contact number $z$ and the packing fraction $\phi$, see grey dots in Fig.~3.
For high packing fractions, the contact number tends to $z=6$, the
expected value for disordered cellular structures [\citet{weairebook}]. 
However, we
observe that, as the foam packing fraction is reduced, the average
contact number decreases, ultimately reaching $z_c=4$, at a packing fraction around $\phi =0.84$. Below this value the foam loses stability.
Both the value of the critical density and contact number at the
unjamming point are fully consistent with predictions.

We also plot the average over all images for each packing fraction (black circles).  Clearly the
variation of $z$ with $\phi$ is similar to a square root, and to
compare our data to predictions we fit our data to a power law fit
of the form ${z} = 4+ z_0*(\phi-\phi_c)^{\beta}$ (red curve in
Fig.~2. The best fit gives us $z_0 =4.02 \pm
0.20$, $\phi_c =0.842 \pm 0.002$ and $\beta=0.50 \pm 0.02$, in
remarkable agreement with theoretical predictions by O'Hern et al
and Durian [\citet{epitome, durianprl95}], who found $z_0 =3.6 \pm 0.5$,
$\phi_c =0.841 \pm 0.002$ and $\beta=0.49 \pm 0.03$.

\begin{figure}[tbh]
\begin{center}
\includegraphics[width=.60\textwidth]{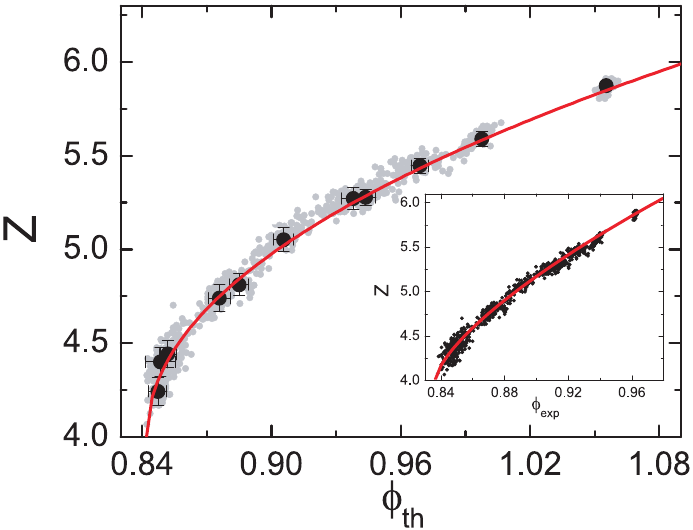}
\caption{ Average contact number versus $\phi$ for experimental bidisperse foams: grey dots indicate data for each individual realization, black circles indicate averages for each globally set packing fraction. Solid red line is fit of the form ${z} = 4+ z_0*(\phi-\phi_c)^{\beta}$, with $z_0 =4.02 \pm
0.20$, $\phi_c =0.842 \pm 0.002$ and $\beta=0.50 \pm 0.02$. Inset is data plotted versus experimentally determined packing fraction
$\phi_{\rm exp}$. The fit has a power law exponent of 0.70.} \label{fig3}
\end{center}
\end{figure}
In simulation studies the packing fraction is
often calculated by counting the overlaps between bubbles {\em twice}.
In the experiment we are restricted to extracting an area fraction from images. We therefore convert our experimentally determined packing fractions to the corresponding theoretical packing fractions. If we do not convert our
experimentally determined packing fractions, we find an effective
scaling exponent $\beta=0.70$ when plotting $z$ as a function of
$\phi_{exp}$ , see inset of Fig.~3.

We were not the the first to experimentally investigate the
scaling of $z$ with $\phi$. Majmudar et al. [\citet{majmudarprl07}] have
extracted the same quantities from images of two-dimensional,
frictional, photoelastic discs and compared these to predictions
from simulations. From their data it appears the prefactor Z$_0
\approx 16$, inconsistent with simulations. Our results do allow
for a direct comparison with predictions for {\em frictionless}
jamming, which can be seen from the excellent agreement between
parameters. We also note that Bruji\`c et al. recently have probed
this squareroot scaling in 3D emulsions (private communication).

We conclude that, as far as static structure is concerned, (2D)
foams and numerical models for jamming are in excellent agreement.
Disorder is crucial for real foams.

\section{Disorder and fluctuations in foam flows: Experiments}

In a series of experiments [\citet{katgertprl,katgertpre09,mobiusepl}] we have
recently probed the role of disorder in the flow of 2D foams. In
these experiments, we shear a foam monolayer that is floating on a soapy solution and bound by a glass plate.  We calculate averaged
velocity profiles and track the bubble displacements. Our
two main findings are 1) that disorder plays a crucial role in
the rheology --- the global rheology is very different than what
you would expect based on measurements of the local drag forces --- and
that 2) fluctuations become increasingly strong as the strain rate is {\em lowered}.

{\bf Drag forces ---} To probe the role of disorder, we have
compared the averaged flow profiles and bubble trajectories of a
ordered, crystalline foam and a disordered, bidisperse foam
(Fig.~4). For monodisperse foams (Fig.~4a), the bubbles move past
each other in a ziplike fashion along the crystal planes of the
hexagonal lattice formed by the bubbles, see Fig.~4b for tracks. The velocity
profiles (Fig.~4b) are strongly localized and are independent of driving
velocity. For disordered foams (Fig~4c)the situation is vastly different:
the bubble motion becomes disordered and organizes into the swirly,
collective patterns typically seen in materials near jamming, see Fig~4d. The velocity
profiles now depend on the driving velocity, and become
increasingly delocalized as the driving velocity is decreased.

\begin{figure}[h!]
\begin{center}
\includegraphics[width= \textwidth]{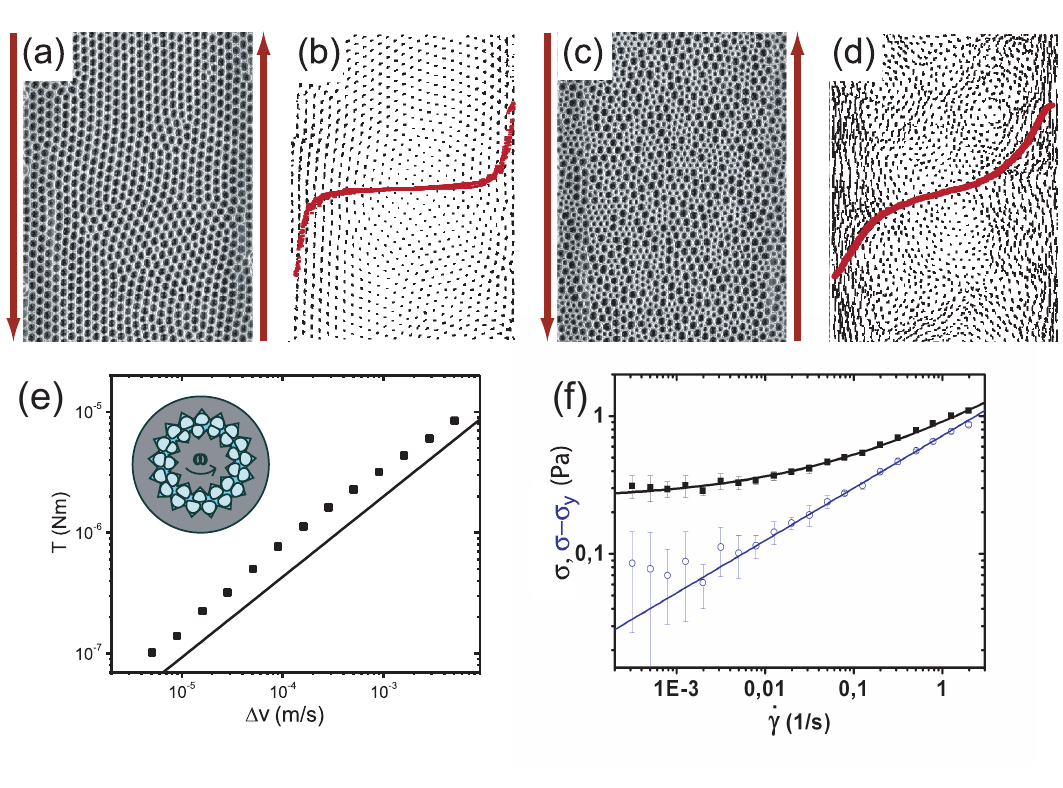}
\caption{(a) Top view of linearly sheared monodisperse foam. Arrows indicate direction of shear at the boundaries. (b) Bubble tracks for monodisperse foam in (a): bubbles slide along planes thus move affinely; the resulting velocity profiles are strongly localized (red curve). (c) Bidisperse foam,  and its corresponding bubble tracks (d): swirly, non-affine motion is clearly visible, and the velocity profile exhibits less localization.  Ingredients into the model describing these flows: (e) Viscous friction between to {\em ordered} bubble rows (see inset) as a function of $\Delta v$. Solid line $\sim \Delta v^{2/3}$. (f) Stress-strain rate relation for {\em disordered} two-dimensional foam: black solid line indicates fit to  the Herschel-Bulkley expression: $\sigma = \sigma_y + A \dot \gamma^\beta$, with $\beta=0.36$. This power law scaling is highlighted by subtracting the yield stress $\sigma_y$ (blue data).}\label{fig4}
\end{center}
\end{figure}

To understand the flow profiles, we introduce a simple force
balance model in which the averaged viscous friction between
bubbles that move past each other, $F_v$, is balanced by the
viscous friction due to the bubbles moving past the glass plate,
$F_{bw}$ [\citet{weaireprl06, katgertprl, katgertpre09}]. We note here that without the top plate, the flow
profiles are essentially linear [\citet{denninpre06}] --- the shear banding seen here is thus simply due to the top plate
[\citet{weaireprl06,schallvanHeckeARFM}]. We check rheometrically that the top plate drag is given by the classic result of Bretherton $F_{bw} \sim v^{2/3}$ [\citet{bretherton}]. We then extract the scaling of interbubble friction  $F_v$
with velocity gradient by fitting the model to the velocity
profiles.

Our model predicts rate independent velocity profiles, as seen in
the ordered foam, only if the interbubble viscous friction scales
in the same way as the bubble-wall friction: $F_{v} \sim 
\Delta v^{2/3}$, where $\Delta v$ is the velocity difference
between neighboring bubbles. We verify this scaling with rheological 
measurements on two ordered bubble layers gliding past each other, 
and find that the bare viscous friction
between bubbles moving past each other indeed scales in exactly
the same way with the velocity difference between bubbles as the
frictional force between bubble and top plate, see Fig.~4e. Hence, the
rate-independence of the observed flow profiles for ordered foams
indicates that the local  viscous drag law immediately and
trivially sets the global drag forces.

For the disordered foams, the remarkable thing is that their rate
dependent profiles are consistent with the force balance model if
the exponents for bubble-bubble and bubble-wall friction are
{\em different}. In fact, we excellently fit all our velocity profiles
by our model with a bubble-wall friction as dictated by
Bretherton's result and an average interbubble friction that
scales as $\Delta v ^{0.36}$. Since the local friction law between
bubbles is independent of the bubble configuration, we conclude
that in this case the translation from local to global drag forces
is highly nontrivial.

We have further checked this remarkable result by bulk rheometry
on disordered two dimensional foam layers. Bulk foam flow curves
are commonly fit with the phenomenological Herschel-Bulkley
constitutive relation, linking the stress $\sigma$ with the strain
rate $\dot \gamma$ :
\begin{equation}\label{HB}
\sigma = \sigma_y + A \dot \gamma^\beta \, 
\end{equation}
with $\sigma_y$ the yield stress. In this measurement (Fig.~4f), the flow exponent $\beta = 0.36$, consistent with our
indirect determination of $\beta$ via the flow profiles. We thus
find that disorder changes the effective viscous friction between
bubbles in a highly nontrivial way: the averaged viscous
dissipation inside the foam is enhanced with respect to the
ordered case, leading to less localized and rate dependent flows.

In the next section, we will present a theoretical model that
fully captures the nontrivial translation from the local exponent
of 2/3 to the global exponent of 0.36.

We also point out here another difference between the measured
rheology of disordered foams and ordered foams: the disordered
foams have a finite yield stress $(\sigma_y)$, while the ordered
foams exhibit pure power law scaling. As we will discuss in section
V.C, this is deeply connected to the relation between fluctuations and energy dissipation in the system.

{\bf Fluctuations ---} We have also studied the bubble
displacements in detail via tracking of the irregular bubble
motion [\citet{mobiusepl}]. After subtracting their average velocity, we have probed
their remaining erratic motion. Since we can probe local strain
rates that span multiple decades, we can plot the mean squared
displacement (MSD) of the bubbles as a function of local strain
rate. For all strain rates we see that the MSD's cross over from
superdiffusive to diffusive behavior at a well defined relaxation
time $t_r$ that corresponds to a MSD of $(0.14 \langle d
\rangle)^2$, which is remarkably similar to the Lindemann
criterion for cage breaking in colloidal suspensions [\citet{besselingprl07}].

When we extract this $t_r$ for different local strain rates, we
observe that it does $not$ scale with the inverse local strain
rate $\dot{\gamma}^{-1}$, which could have expected expected since both 
rearrangements and fluctuations are entirely shear induced. In
contrast, we find that $t_r \sim \dot{\gamma}^{-0.66 \pm 0.05}$.
This result implies that fluctuations $increase$ for slower flows,
in contrast to the commonly accepted viewpoint that foam flows
tend to quasistaticity when lowering the strain rate, reflected in
the expected scaling $t_r \sim \dot{\gamma}^{-1}$. This result can
be made more explicit by replotting the MSD curves as function of
the accumulated strain --- the lower the strain rate, the larger
the MSD at given strain [\citet{mobiusepl}]. Hence fluctuations become stronger the
slower the flow.

\section{Disorder and fluctuations in foam flows: Simulations and Theory}

In this section we will present a jamming-inspired, theoretical
perspective on the flow of foams. For completeness, we first
introduce several variants of the ``bubble model'', a microscopic
model suitable to simulate the flow of wet foams --- readers
familiar with, or not interested in the details of these models
can skip this section. We then discuss the main rheological
features that these models display in direct numerical
simulations. We then show that a basic energy balance equation
implies that the relative strength of fluctuations {\em grows} for
decreasing flow rates, and actually {\em diverges} in the limit of
zero flow rate - consistent with the experimentally observed
growth of the diffusivity discussed above. We finally outline the
main contours of a model for the rheology of soft materials, such
as foams, near jamming. This model combines recent insights on
elasticity near jamming with our observations on the nature of the
fluctuations, and predicts several scaling regimes that should be,
and partly have been, observed in the flow of foams.

\subsection{Computer Models for Foam Flow}
There are various versions of microscopic models suitable for
simulating the flow of wet foams (and other disordered media). All
version we will discuss here stem from the ``bubble model''
introduced in 1995 by Durian [\citet{durianprl95}]. In these models,
all bubbles are represented by soft spherical (disk-like in 2D)
particles, that only interact when in contact. Following Durian,
it is typical to take the equations of motion to be overdamped
[\citet{durianprl95,durianpre97,ono, tewari,olssonprl07,tigheprl10}], an appropriate approximation for slow flows in which damping
dominates inertia. This formally corresponds to the limit in which
bubble masses are set to zero, and hence at any instant the
elastic and viscous forces on each bubble balance. Other authors
elect to retain inertial terms in the equations of motion
[\citet{hatano08,langlois09,otsuki09}]; in the underdamped limit this
produces a qualitatively different rheology.

\begin{figure}[tbh]
\begin{center}
\includegraphics[width=.7\textwidth]{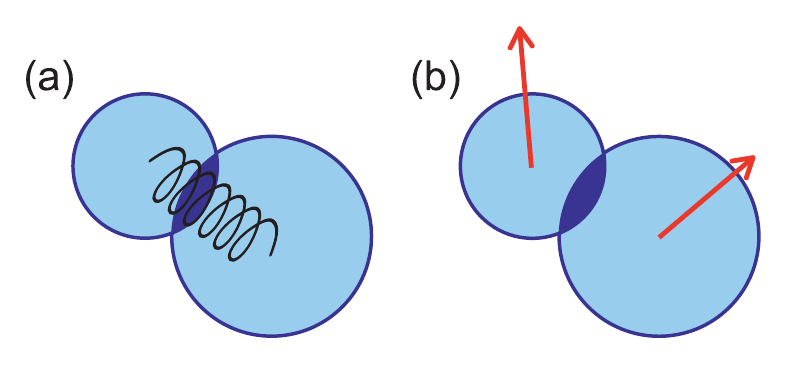}
\caption{(a) Elastic  and (b) viscous forces between bubbles in the bubble model: when bubbles overlap, the strength of their mutual repulsion is a function of the overlap ---  the bubbles act as one-sided springs. The viscous forces are taken to be a function of the velocity difference between bubbles sliding past each other.} \label{fig5  }
\end{center}
\end{figure}

To specify a particular variant of the bubble model one must
describe the elastic and viscous forces between bubbles. Elastic
forces are typically taken to be generated by harmonic ``one-sided
springs'', see Fig.~5a. These springs have rest lengths equal to the sum of the
contacting bubbles' radii; their one-sidedness refers to the fact
that they apply a force only when the spring is compressed,
thereby guaranteeing that the elastic forces are purely repulsive.
Another common force law is the Hertzian interaction, which
describes emulsions rather than foams [\citet{lacasseprl96}]. A ``Hertzian spring'' is
also one sided, but its repulsive force grows as the $3/2$ power
of the spring's compression. In general, then,
\begin{equation}
f_{\rm el} \sim \delta^{\alpha_{\rm el}} \,,
\end{equation}
where $\delta$ is the dimensionless compression of the spring and
$\alpha_{\rm el} = 1$ ($3/2$) corresponds to harmonic (Hertzian)
forces --- for the simulations we describe below $\alpha_{\rm el}
= 1$.

Similarly, it is natural to invoke a viscous force law  in which
the relative velocity of contacting bubbles is damped (Fig~5b):
\begin{eqnarray}
f_{\rm visc}^\parallel &=&
 b^\parallel (\Delta v^{\parallel})^{\alpha_{\rm visc}}  \nonumber \\
f_{\rm visc}^\perp &=& b^\perp (\Delta v^{\perp})^{\alpha_{\rm
visc}} \,. \label{eqn:fullvisc}
\end{eqnarray}
For the simulations we describe below $\alpha_{\rm visc} = 1$,
which corresponds to linear viscous dissipation. Real foams,
however, are believed to have $\alpha_{\rm visc} < 1$, with the
actual value depending on the surfactant used [\citet{denkovsm09}]. For the mobile
surfactants used in the experiments described above, our
measurements indicate $\alpha_{\rm visc} = 2/3$, see Fig.~4(e). We shall see that changes in the microscopic exponent
$\alpha_{\rm visc}$ have measureable ramifications for the
macroscopic rheology. We also take $b^{\parallel} = b^\perp$, so
that the dissipation is indifferent to whether the particles are
moving together, moving apart, or sliding. Other authors have
taken $b^\perp = 0$, so that sliding does not
dissipate [\citet{hatano, otsuki09}], though this case may be pathological.

Finally we point out that there is another, qualitatively
different, viscous force law commonly used in the literature
[\citet{durianprl95,durianpre97,ono,tewari,olssonprl07}]. Known as ``mean
field'' dissipation, it specifies that each bubble experiences a
damping force proportional to the difference between its
instantaneous velocity and the {\em mean} velocity $ \vec v_{\rm
lin} = \dot \gamma x \, \hat y$ of a bubble at its position:
\begin{equation}
\vec f_{\rm MF} = -b_{\rm MF}(\vec v - \vec v_{\rm
lin} )
\,. \label{eqn:MF}
\end{equation}
This expression assumes the linear velocity profile established by
Lees-Edwards boundary conditions. It sacrifices more realistic
modeling for numerical convenience: in overdamped dynamics,
Eq.~(\ref{eqn:MF}) can be simulated far more efficiently than
Eq.~(\ref{eqn:fullvisc}).

\subsection{Phenomenology} In the static limit, all microscopic models introduced above reduce to the simple
models that are studied to probe elastic properties near the
jamming transition. Most crucial, in order for the material to
have a finite rigidity, they need to be packed at a sufficiently
large packing fraction --- the critical packing fraction
corresponds to extremely wet foams where the bubbles just touch.
Once such packings are created, they are subjected to a fixed
strain rate and the shear stress can be measured numerically ---
in some cases, the stress has been fixed and the strain rate has
been measured [\citet{olssonprl07}], but in all cases, flow profiles
have been found to be simply linear (i.e., no shear banding ---
when we include additional drag terms, modeling interactions with
a top plate, the same shear banding that is seen experimentally is
recovered).

There are certain generic features of the macroscopic rheology
that prevail for $\phi > \phi_c$ in simulations of any version of
the bubble model discussed above. Namely, the constitutive
relation is qualitatively similar to the Herschel-Bulkley
constitutive relation defined in Eq.~(\ref{HB}) \footnote{We note here that we do not
believe that the flow curves are exactly of this form, although
this is almost always a good approximation}. The exponent
$\beta$ is typically less than one, a property known as ``shear
thinning.''  In numerics one observes that $\sigma_y$ varies with
$\phi$ and vanishes at $\phi_c$, which motivates us to write
$\sigma_y \sim \Delta \phi^\Delta$ for some $\Delta$. Note that
because $\sigma = \sigma_y$ in the limit $\dot \gamma \rightarrow
0$, it is natural to expect that the exponent $\Delta$ depends on
$\alpha_{\rm el}$ but {\em not} on $\alpha_{\rm visc}$, a
dynamical quantity. By contrast $\beta$ is a dynamical exponent
which will turn out to depend on both $\alpha_{\rm visc}$ and $\alpha_{\rm el}$. The
challenge is to identify how $\Delta$, $A$, and $\beta$ depend on
the microscopic details of the model.

\subsection{Fluctuations}

Slow foam flows are dominated by fluctuations. For example it is
widely believed that the main contribution to dissipation in dry
foams stems from localized neighbor-swapping, or T1, events
[\citet{hohlerreview}]. Here we focus on the wet limit, where T1 events are
not well defined. Nevertheless, here again there is a precise relation
between the fluctuations and dissipation.
{\bf Power balance ---} If one shears a foam (or any other system)
at a rate $\gammadot$, and it resists this flow with a shear
stress $\sigma$, the product of the two gives the average rate of
work done on the system. This work has to be dissipated, and for
wet foams the only dissipative mechanism is the relative motion of
neighboring bubbles. Let us first take a simple model, where the
viscous force $f_{visc}$ between bubbles is proportional to their
relative velocity $\Delta v$. In a time interval $\Delta t$, the
amount of energy dissipated per pair of sliding bubbles is then
$f_{visc} \times \Delta t \times \Delta v$, which scales as
$\Delta v^2 \Delta t$. The crucial observation is that the energy fed into the system by shear needs to be
balanced by the amount of energy dissipated by local bubble
sliding:
\begin{equation}\label{balance}
\sigma \gammadot  \sim \left< \Delta v^2\right>~,
\end{equation}
where the brackets denote averaging over the system \footnote{In
all scaling arguments that will follow we focus on the typical
scale of quantities, and ignore correlations --- we do not believe
these will change our results in an essential manner.}.

We now show the powerful consequences of this simple relation.
Let us first imagine that our foam behaves as a Newtonian fluid,
for which $\sigma = \eta \dot \gamma$. In that case the steady
state power balance equation Eq.~(\ref{balance}) requires that the
$ \eta \gammadot^2 \sim\left< \Delta v^2\right>$, so that $|\Delta
v|$ and $\gammadot$ scale in the same way.

However, foams do not behave as Newtonian fluids, particularly for slow flows. 
Let us assume that the stress varies as a power law of the strain rate, 
which is a good approximation of a typical Herschel-Bulkley rheology at high flow
rate: $\sigma \sim \gammadot^\beta$. Substituting this into the
steady state power balance equation, Eq.~(\ref{balance}), one
obtains $\gammadot^{1+\beta} \sim\left< \Delta v^2\right>$, so
that $|\Delta v|$ and $\gammadot$ {\em do not} scale in the same
way; rather $|\Delta v| \sim \gammadot^{(1+\beta)/2}$. For the
typical case that $\beta < 1$, one concludes that the velocity
fluctuations decay sublinearly with strain rate, or in other
words, that the {\em relative} velocity fluctuations, $|\Delta
v|/\gammadot$ diverge as $ \gammadot^{(\beta-1)/2}$. This
divergence is even stronger for slow flows for which the stress
reaches a plateau value: $\sigma \approx \sigma_y$. The steady state power
balance equation Eq.~(\ref{balance}) then predicts $|\Delta
v|/\gammadot$ to diverge as $ \gammadot^{-1/2}$.

The divergence of the fluctuations suggests that there is no
well-defined quasistatic limit, at least not as far as the
trajectories of the bubbles are concerned: the slower you go, the
bigger the fluctuations. Imagine you are given two movies of foam
flows at two different flow rates --- but you are not told which
one is the fastest, nor do you know the frame rate of these
movies. By adjusting their playback speed, you can make the
average flow rate of the two movies equal
--- the energy balance argument predicts that the
amount of fluctuations would be largest in the movie corresponding
to the slowest flow. We stress here that, at least qualitatively,
this is precisely what we observed in the experiments on diffusion
of foam bubbles in flow.

{\bf Characteristic Scales ---} The fluctuations introduce a
nontrivial dynamical time scale into the problem: $t_{\rm dyn} =
d/|\Delta v|$, where $d$ is a typical bubble diameter --- this
time can be thought of as characterizing the time scale over which
local rearrangements take place. In the simplest case of a
Newtonian fluid, for which $|\Delta v| \sim \gammadot$, $t_{\rm
dyn}$ is nothing more than the inverse strain rate.

For non-Newtonian fluids, however, $\Delta v$ scales nontrivially, hence the dynamical timescale does, as well. For a pure power law fluid we
find $t_{\rm dyn} \sim \gammadot^{(1+\beta)/2}$, and for the yield
stress case we find $t_{\rm dyn} \sim \gammadot^{+1/2}$.

One can translate this time scale into a characteristic strain
scale necessary to induce rearrangements: $\gamma_{\rm dyn} =
t_{\rm dyn} \gammadot$. This characteristic strain thus vanishes
for slow flows --- the slower the flow, the smaller the overall
strains necessary to induce substantial rearrangements.

{\bf Relation between local viscous drag, fluctuations and
rheology ---} Let us now clarify the relations between local
dissipation, fluctuations and global rheology. First, let us
generalize the viscous force to be nonlinear: $f_{\rm visc}
\propto |\Delta v|^\alpha$. Now assume that the macroscopic
rheology is of the form $\sigma \propto \dot \gamma^\beta$. The
power balance equation then reads $\gammadot^{\beta+1} \sim \left<
\Delta v^{1+\alpha}\right>$, which will give nontrivial
fluctuations whenever $\alpha \neq \beta$.

This shows that details matter in the sense that different local
dissipative laws directly affect the precise form of the balance
Eq.~(\ref{balance}) --- in this sense, the physics is not
universal. On the other hand, as long as $\beta < \alpha$, the
fluctuations become stronger for slower flow. We will see that
this inequality is in general satisfied, so that the divergence of
fluctuations is robust.

The balance arguments can also be used to rationalize part of our
experimental findings. Recall that in rheological experiments in
which two ordered rows of bubbles were slid past each other with
velocity $\Delta v$ [\citet{katgertprl, katgertpre09}], see Fig.~4e, the time-averaged shear stress
$\sigma$ was found to scale as a simple powerlaw: $\sigma \sim
\Delta v^{\alpha}$. In particular, there was {\em no} force plateau at
low $\Delta v$. In contrast, for a disordered bubble raft, the stress had a
Hershel-Bulkley form with a plateau for low strain rates and a
power law different from $\alpha$: $\sigma \sim \sigma_0+ A
\gammadot^{\beta}$, see Fig.~4f.

The finite plateau we can understand as follows. As we have seen,
the energy balance argument shows that for flows in the regime where
the stress is on a plateau, the relative fluctuations must diverge
when the strain rate goes to zero. Since the fluctuations are
constrained in the ordered system, they cannot diverge, energy
cannot be dissipated strongly enough, and as a result {\em there cannot be a
yield stress} --- consistent with our findings. In the disordered
system, the fluctuations are not constrained, and nothing forbids
the emergence of a finite yield stress.

We stress here that the role of disorder is to facilitate the
nontrivial role of the fluctuations. In ordered systems, such as
the ordered foams we discussed above, whole rows of bubbles slide
past each other, and there is simply not enough freedom to allow
for large fluctuations. In disordered systems, the bubbles have
much more freedom to choose their path. For the case of elasticity
of disordered media, discussed in section II, disordered bubble
motion was intimately connected to anomalous scaling of the shear
modulus. Here we see that for the flow of disordered media, nontrivial fluctuations are connected to anomalous scaling of the
stress-strain rate relation.

The difference between the local and global rheological exponents
$\alpha$ and $\beta$ is also intimately related to the fact that
the fluctuations are rate dependent. But the single balance
equation Eq.~(\ref{balance}) is not sufficient to predict both
$\beta$ and the fluctuations --- we need additional arguments to
obtain a set of closed equations. In the next section we will
introduce such equations, and produce a definite prediction of
$\beta$ as function of the local drag force $\alpha$.

\subsection{Scaling model for foam flows}

We now introduce a model that seeks to explain bubble model
rheology for $\phi> \phi_c$. Our goal is to give the flavor of the
model, without delving too far into details.  For simplicity, we first
fix the elastic ($\alpha_{\rm el}$) and viscous exponent
($\alpha_{\rm visc}$) to one.

The scaling model has three ingredients. These are
\begin{enumerate}
\item the system is in power balance
\item a flowing foam can be mapped to a static system that has been sheared
through an effective strain $\gamma_{\rm eff}$
\item the stress as a function of $\gamma_{\rm eff}$ is given by the constitutive
relation for sheared disordered spring networks
\end{enumerate}
One of the difficulties in explaining this model is that there are
several different regimes with different forms for the effective
strain and stress-strain relation. We first will describe the
model in two of the simplest limits, and then briefly point out
how all these results can be generalized.

{\em Critical Regime ---} In the limit of very wet foams, i.e.,
$\phi \rightarrow \phi_c$, the model is particularly simple. We
postulate that the effective strain in the system then simply
equals the dynamical strain introduced above:
$\gamma_{\rm dyn}  \sim \gammadot /|\Delta v|$. For the equation
for the stress, we build on recent work that shows that near
jamming, the linear shear modulus vanishes and the stress scales as $\sigma \sim |\gamma|\gamma$. [\citet{wyart08}].
Now we have a closed set of three equations (we drop all absolute values and
averages):
\begin{eqnarray}
\sigma \gammadot &\sim & \Delta v^2~. \\
\gamma_{\rm eff} &\sim& \gammadot /\Delta v~. \\
\sigma &\sim& \gamma_{\rm eff}^2~.
\end{eqnarray}
We have identified $\gamma$ with $\gamma_{\rm eff}$; this is point 3 in the above list.
These equations can easily be solved and yield a prediction for
the rheology of the form $\sigma \sim  \gammadot^{1/2}$, which is
in very good agreement with numerical simulations of the bubble
model [\citet{tigheprl10}].

{\em Yield Stress Regime ---} In the limit of very slow flows and
for $\phi >  \phi_c$, the model also becomes very simple. In that
regime, the effective strain is no longer expected to be dominated
by dynamic effects, and we postulate that the effective strain in
the system then equals $\Delta \phi := \phi -  \phi_c$ [\citet{tigheprl10}]. For
the equation for the stress, recent work shows that away
from jamming, the shear modulus $G$ scales with the distance to
jamming as  $G \sim \Delta z \sim \sqrt{\Delta \phi}$. In
summary:
\begin{eqnarray}
\sigma \gammadot &\sim & \Delta v^2~. \\
\gamma_{\rm eff} &\sim& \Delta \phi. \\
\sigma &\sim& \sqrt{\Delta \phi}\gamma_{\rm eff}~.
\end{eqnarray}
These equations can easily be solved, and yield as prediction that
the stress is a constant $\propto \Delta \phi^{3/2}$.

{\em Transition Regime ---} The general form of the model is
obtained by combining these regimes, so that $\gamma_{\rm eff} =
a_1  \Delta \phi + a_2 \gammadot /\Delta v$, and $\sigma = a_3
\sqrt{\Delta \phi}\gamma_{\rm eff} + a_4 \gamma_{\rm eff}^2$,
where $a_i$ are numerical constants to be determined. It turns out
that there are three different regimes, and in all cases, either
the first or the second term in these sums dominates \footnote{In
principle there are four regimes, but the combination of terms in
this fourth regime never dominate the physics --- see [\citet{tigheprl10}].}.
The third regime we refer to as the transition regime, and here
the three equations are:
\begin{eqnarray}
\sigma \gammadot &\sim & \Delta v^2~. \\
\gamma_{\rm eff} &\sim& \gammadot /\Delta v~. \\
\sigma &\sim& \sqrt{\Delta \phi}\gamma_{\rm eff}~,
\end{eqnarray}
which yields that the stress scales as $\sigma \sim (\Delta
\phi)^{1/3} \gammadot^{1/3}$ --- this third regime came as a
complete surprise, and its existence implies deviation from the
usual Herschel-Bulkley phenomenology. This deviation is difficult to
observe in numerics but consistent with our observation that very
often, Herschel-Bulkley-fits to experimental or numerical data underestimate the data in the crossover regime between yield
stress and power law behavior - precisely the regime where the
transition rheology is predicted.

By substituting the various solutions we have obtained into the
general expressions for the effective strain and stress, one can
do a self consistency check to see for which strain rates, which
regime should dominate --- we find that the Yield Stress regime
dominates for $\gammadot \lesssim \Delta \phi^{7/2}$, the Critical Regime
for $\gammadot \gtrsim \Delta \phi^{2}$ and the Transition regime in the
range in between --- which can span arbitrarily many decades when
$\Delta \phi$ tends to zero.

{\em General Microscopics ---} What happens when the microscopic
exponents ($\alpha_{\rm el}$) and ($\alpha_{\rm visc}$) are
unequal to one?

For the Yield Stress regime, the stress-strain relation is
affected by the value of $\alpha_{\rm el}$, as it sets the scaling
of the shear modulus $G$ with $\Delta \phi$. Prior work has shown
that for nonlinear interactions, the stress strain relation
becomes $\sigma \sim \Delta \phi^{\alpha_{\rm el}} - 1/2$
[\citet{epitome}]. Thus the yield stress varies with $\alpha_{\rm
el}$ according to $\Delta = \alpha_{\rm el}+1/2$. For $\alpha_{\rm
el} = 3/2$, believed to describe dense emulsions and microgel
suspensions [\citet{lacasseprl96, nordstrompre10}], this gives $\sigma_y \sim \Delta \phi^2$, in good
agreement with experimental measurements by \citet{masonjcis96} and \citet{nordstromprl10}, which both
find $\Delta \approx 2$.

For the Critical Regime, both the elastic and viscous exponent
matter. Let us keep the elastic exponent equal to one, which is a
good approximation for foams. Recall that $\beta$ should, in
general, depend on $\alpha_{\rm visc}$. Repeating the analysis in
the critical regime for arbitrary $\alpha_{\rm visc}$ yields
\begin{equation}
\beta = \frac{2\alpha_{\rm visc}}{\alpha_{\rm visc}+3} \,.
\end{equation}
For $\alpha_{\rm visc} = 2/3$, as is the case for mobile
surfactants, we obtain $\beta = 4/11 \approx 0.36$, in remarkably
good agreement with the experimentally determined value $\beta
\approx 0.36$ in \citet{katgertprl}.

Finally, we point out here that for dense granular media, the main
dissipation comes from sliding friction, which could be seen as a
viscous interaction with exponent zero --- the frictional forces
do not change appreciably with the sliding rate. In granular
media, it has been taken as a triviality that both the local and
global interactions are very similar, namely frictional. We doubt
that granular flows are affine, but we believe that this
correspondence between local and global flow behavior is  a lucky
coincidence - we note here that for $\alpha_{\rm visc}=0$, the
global flow exponent $\beta $ also becomes zero in our model.

\subsection{Critical collapse}

\begin{figure}[h!]
\begin{center}
\includegraphics[width= 0.6\textwidth]{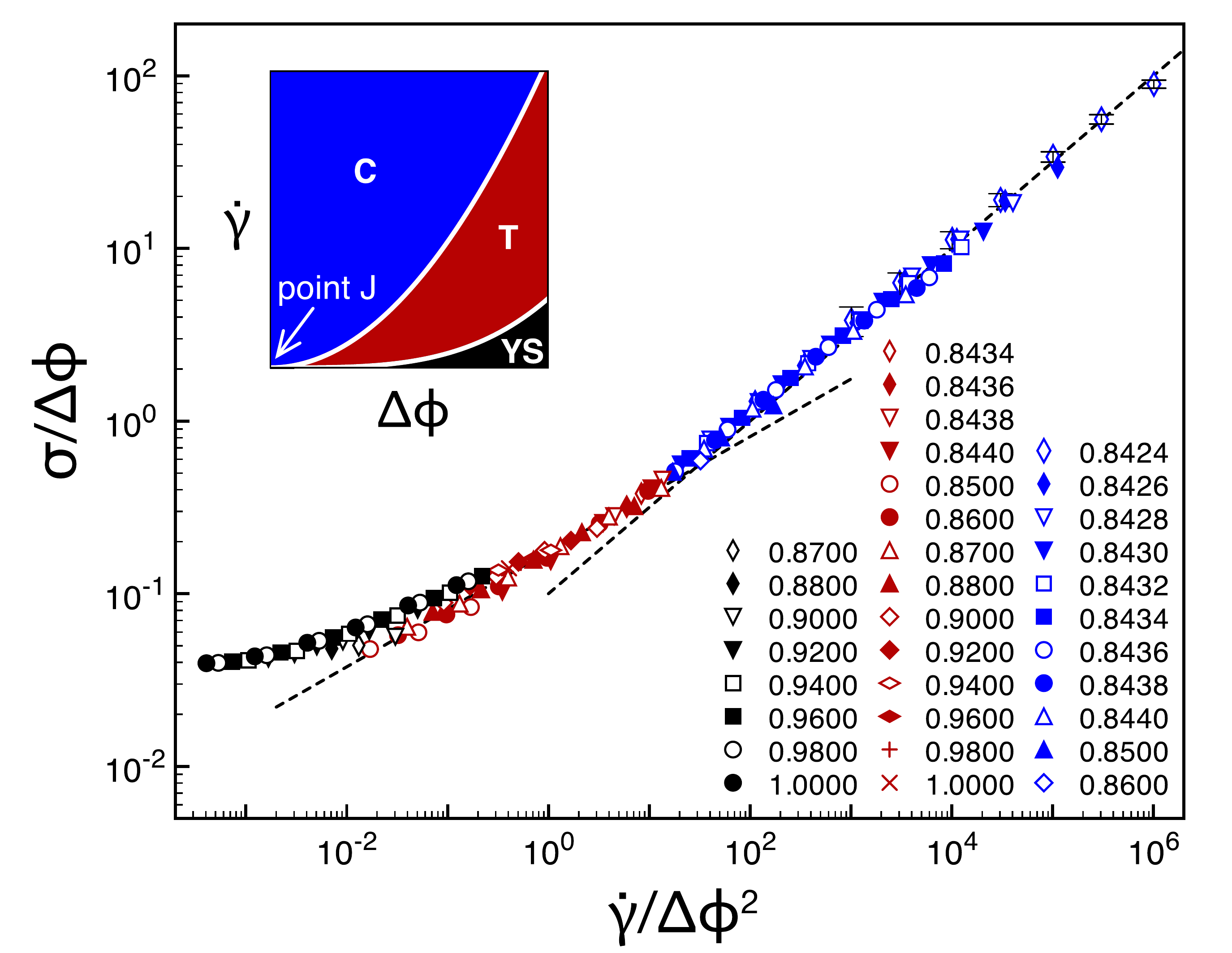}
\caption{Scaling collapse of flow curves obtained from bubble model simulations as described in Section 5.1. Flow curves are obtained for 4 decades in strain rate and 3 decades in $\Delta$, see legend. Rescaled coordinates are $\sigma/\Delta \phi$ and $\dot{\gamma}/\Delta \phi^2$, appropriate for parameters spanning the Transition and Critical
regimes. Dashed lines are guides to the eye with
slopes 1/3 and 1/2. Inset: Boundaries between the yield stress,
transition, and critical regimes in the $\Delta \phi - \dot{\gamma}$ plane.}
\label{fig6}
\end{center}
\end{figure}
A powerful test of any critical scaling prediction is to plot numerical or experimental data in rescaled coordinates and look for collapse to a master curve. The quality of collapse is a strong test of theory, and the master curve itself is a revealing depiction of the underlying physics. The rheological model described above indeed predicts such master curves, but there is some subtlety owing to the existence of not one but two crossover strain rates.

To illustrate the subtlety, let us first consider an example with only one crossover. The Herschel-Bulkley flow rule contains two regimes, a yield stress regime $\sigma \approx \sigma_{\rm y} \sim \Delta \phi^\Delta$ and a critical regime $\sigma \sim \dot \gamma^\beta$. The crossover between the two regimes occurs at the strain rate $\dot \gamma^*$ for which the two regimes are comparable: $\dot \gamma^*  \sim \Delta \phi^{\Delta /\beta}$. Dividing by $\Delta \phi^\Delta$, the Herschel-Bulkley flow rule can be rewritten
\begin{equation}
\frac{\sigma}{\Delta \phi^\Delta} = {\rm const} + A \left(\frac{\dot \gamma}{\Delta \phi^{\Delta/\beta}} \right)^\beta \,
\end{equation}
provided $A$ does not depend on $\Delta \phi$ as well [\citet{katgertpre09}].
Note that the term in parentheses is proportional to $\dot \gamma / \dot \gamma^*$, and that all the dependence on $\Delta \phi$ is contained in the ratios $\sigma/\Delta \phi^\Delta$ and $\dot \gamma/\Delta\phi^{\Delta/\beta}$. Therefore {\em if} the rheology is described by a Herschel-Bulkley rule (or another flow rule with a single crossover rate), plotting the rescaled stress and strain rate,
\begin{equation}
\tilde \sigma = \frac{\sigma}{\Delta \phi^\Delta} \,\,\, {\rm and} \,\,\,
\dot{\tilde \gamma} = \frac{\dot \gamma}{\Delta \phi^{\Delta/\beta}} \,,
\end{equation}
should produce collapse to a master curve of the form ${\rm const} + A x^\beta$. Alternatively, if the exponents $\Delta$ and $\beta$ are unknown, such a scaling plot can be a way to determine them: one looks for the values of $\Delta$, $\beta$, and $\phi_c$ that produce the best data collapse.

Such scaling plots can be made for any curve characterized by two qualitatively different regimes: the first step is always to identify the crossover strain rate (or whatever quantity is on the $x$-coordinate). If a curve has more than two regimes, and hence more than one crossover, one generally cannot collapse all data by resaling with the single scaling parameter $\Delta \phi$. The flow model from above predicts not two but three scaling regimes, hence, according to the model, one should not expect to be able to make a scaling plot that collapses all data near $\phi_c$ for a broad range of $\dot \gamma$. One can, however, still collapse data from any two adjacent regimes. Consider the Transition and Critical regimes for $\alpha_{\rm el} = 1$ and $\alpha_{\rm visc} = 1$, for which $\sigma \sim \Delta \phi^{1/3}\dot \gamma^{1/3}$ and $\sigma \sim \dot \gamma^{1/2}$, respectively. The crossover rate is $\dot \gamma^* \sim \Delta \phi^2$, and data from these two regimes should collapse for the rescaled coordinates
\begin{equation}
\tilde \sigma = \frac{\sigma}{\Delta \phi} \,\,\, {\rm and} \,\,\,
\dot{\tilde \gamma} = \frac{\dot \gamma}{\Delta \phi^{2}} \,.
\end{equation}
This is shown in Fig.~\ref{fig6} and the collapse is indeed very good. Similarly, data from the yield stress and transition regimes can be collapsed by plotting 
\begin{equation}
\tilde \sigma = \frac{\sigma}{\Delta \phi^{3/2}} \,\,\, {\rm and} \,\,\,
\dot{\tilde \gamma} = \frac{\dot \gamma}{\Delta \phi^{7/2}} \,.
\end{equation}
Note, however, that generating data in the yield stress regime is numerically challenging; Fig.~\ref{fig6}, for example, does not have enough data points in this regime to allow for a convincing test of the model.

\begin{figure}[h!]
\begin{center}
\includegraphics[width= 0.6\textwidth]{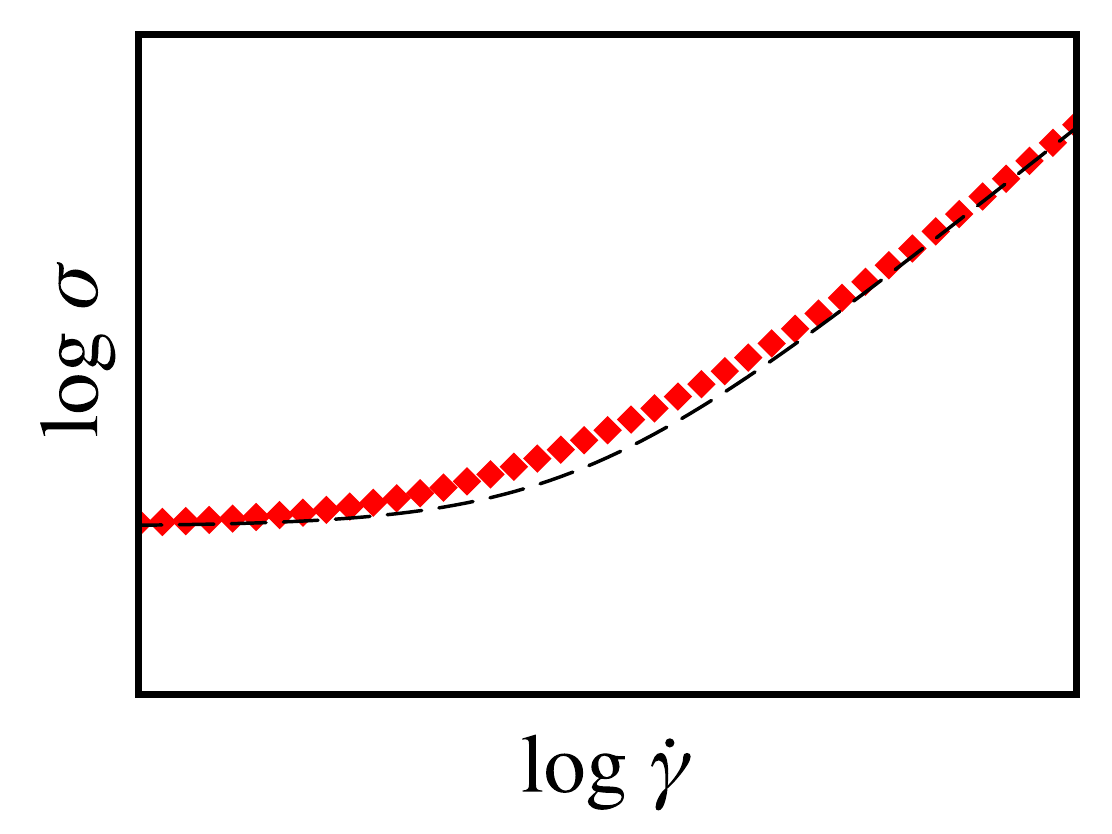}
\caption{Numerical solution to our scaling model ($\blacklozenge$) and Herschel-Bulkley model with the same asymptotic behaviour (dashed line) compared. The Herschel-Bulkley model underestimates the cross-over as it ignores the Transitional regime.}
\label{fig7}
\end{center}
\end{figure}

It is interesting to ask what happens if we attempt to make a scaling plot that assumes a single crossover, as in the Herschel-Bulkley flow rule, when in fact there are multiple crossovers. In effect, this ignores the transition regime, and one should find that data from the transition regime fails to collapse. Fig.~\ref{fig7} shows a plot of data generated by numerically solving our scaling model ($\alpha_{\rm el} = \alpha_{\rm visc}=1$). The dashed line indicates a Herschel-Bulkley law: $\sigma_y + A \dot \gamma^{1/2}$, with $\sigma_y$ and $A$ chosen so as to match the asymptotics of the model. Note that the curve has a soft elbow: the crossover is much slower than the Herschel-Bulkley form would predict. While real experimental data likely are noisy enough to mask the poor collapse in this regime, the soft elbow does appear to be a feature of real flow curves: fits to the HB flow rule typically fare poorly near the crossover point and underestimate the stress, consistent with the existence of a transition regime.

\subsection{Conclusion and Outlook}

In this paper, we have proposed that static and dynamic properties
of foams can be captured by recent ideas stemming from jamming -
and we have also illustrated how some of these theoretical ideas
grew out of experimental questions ("how does a foam lose
rigidity") and observations ("fluctuations in foams seem to grow").

We suggest two broad directions for future research. On the one
hand, there is a general lack in experimental data confronting
some of the predictions, for example in the degree of nonaffinity
as a function of wetness [\citet{jammingreview}], or the distribution of
relative velocity fluctuations as function of strain rate [\citet{tigheprl10}]; more experiments are called for. 
In particular there are only a handful of studies of the mechanics and flow of very wet foams
or emulsions - the recent advent of combined
3D confocal imaging and rheology of emulsions [\citet{dinsmore, brujic, bonn}] should open new
experimental avenues to probe the critical regime.

On the other hand, many of the commonly observed phenomena in soft
materials are without a sound description. Here we have focused
on steady state rheology, but in practice, oscillatory rheology is
the preferred experimental tool to capture the general
visco-elastic behavior of soft materials. One of us, has
recently developed a theoretical description of linear oscillatory
rheology [\citet{tigheprl11}], but there are many other open
questions. What is the physics of plastic rearrangements? What is
the fate of T1 evens when the foam gets increasingly wet? What
about memory effects and reversibility [\citet{dennin08}]?

We look forward to many more surprises in the rich physics of
collections of bubbles and droplets.

\end{document}